\documentclass[twocolumn,letterpaper,10pt]{article}
\usepackage{iasted}
\usepackage{times}
\usepackage[dvips]{graphicx}
\usepackage{amssymb,amsmath}

\begin{document}

\date{}

\title{AN EXPLORATION OF OPENCL ON MULTIPLE HARDWARE PLATFORMS FOR A NUMERICAL RELATIVITY APPLICATION}

\author{
Niket K. Choudhary and Sandeep Navada\\
Department of Electrical and Computer Engineering,\\
North Carolina State University, \\
Raleigh, NC 27695. \\
\and 
Rakesh Ginjupalli and Gaurav Khanna\\
Physics Department,\\
University of Massachusetts Dartmouth,\\
North Dartmouth, MA 02747.\\ 
}

\maketitle

\thispagestyle{empty}

\noindent
{\bf\normalsize ABSTRACT}\newline
{Currently there is considerable interest in making use of many-core processor architectures,
such as Nvidia and AMD graphics processing units (GPUs) for scientific computing. In this  
work we explore the use of the Open Computing Language (OpenCL) for a typical Numerical 
Relativity application: a time-domain Teukolsky equation solver (a linear, hyperbolic, partial 
differential equation solver using finite-differencing). 
OpenCL is the only vendor-agnostic and multi-platform parallel computing framework that has 
been adopted by all major processor vendors. Therefore, it allows us to write {\em portable} 
source-code and run it on a wide variety of compute hardware and perform meaningful comparisons. 
The outcome of our experimentation suggests that it is relatively straightforward to obtain 
{\em order-of-magnitude} gains in overall application performance by making use of many-core 
GPUs over multi-core CPUs and this fact is largely {\em independent} of the specific hardware 
architecture and vendor. We also observe that a {\em single} high-end GPU can match the performance 
of a small-sized, message-passing based CPU cluster.} \vspace{2ex}
   
\noindent
{\bf\normalsize KEY WORDS}\newline
{GPU, OpenCL, Relativity, Astrophysics}

\section{Introduction}
\label{}
Computational scientists and engineers have begun making use of many-core GPU architectures because 
these can provide significant gains in the overall performance of many numerical simulations at a relatively 
low cost. However, to the average computational scientist, these GPUs usually employ a rather unfamiliar 
and specialized programming model that often requires advanced knowledge of their architecture. In addition, 
these typically have their own vendor- and platform- specific software development frameworks (SDKs), that are different 
from the others in significant ways. For example: Nvidia's GPUs use CUDA SDK~\cite{cuda}, AMD's GPUs use Stream 
SDK~\cite{stream}, the STI Cell BE uses IBM Cell SDK~\cite{cell}, while traditional multi-core processors (from 
Intel, AMD, IBM) typically employ an OpenMP-based parallel programming model~\cite{openmp}. Therefore, the average 
computational scientist, with limited time and resources available to spend on deeply specialized software 
engineering aspects of these modern processors, is simply unable to embrace and make effective use of these current 
technologies for advancing science.

In 2009, an open standard was proposed by Apple to bring the software development for all these different 
processor architectures under a single standard -- the Open Computing Language (OpenCL)~\cite{opencl} -- and all 
major multi-core processor and GPU vendors (Nvidia, AMD, IBM, Intel) have adopted this standard for their current and 
future hardware. 

The scientific application that we concentrate on in this work is an application from our Numerical Relativity 
(NR) community -- a time-domain Teukolsky equation solver, which is an explicit, finite-difference, linear, hyperbolic 
(2+1)D PDE solver that uses the 2nd-order Lax-Wendroff numerical evolution scheme~\cite{klpa,mystuff6,mystuff7}. Its 
worth noting that such a solver is in fairly common use, not only in our NR community but also in various other fields of science and 
engineering (for example, in Engineering Electromagnetics, Optics, Acoustics, Fluid Dynamics and other applications of the 
wave equation); therefore we expect that our work would be of strong interest to the larger community of computational scientists. 
It is also worth pointing out that while some NR works~\cite{others1,others2,others3} in the recent past have investigated 
the use of GPUs for finite-difference PDE solutions, these have mainly concentrated on making use of vendor supplied frameworks 
and libraries (such as Nvidia's CUDA) and achieving maximal performance. In our work here, we concentrate entirely on the OpenCL framework 
and pay serious attention to code {\em portability} in addition to performance. In other words, we run identical OpenCL source-code  
on {\em all} the hardware we consider in this work (both CPUs and GPUs). This work is amongst the first\footnote{It is worth clarifying that 
earlier this year, we published a research article~\cite{justin2} that used OpenCL to parallelize our EMRI Teukolsky Code on an
Nvidia Tesla GPU and the Cell BE. However, that particular code is critically different from the one under consideration in the 
sense that it additionally has a very mathematically complex source-term i.e. it is an {\em inhomogeneous} hyperbolic PDE solver; and 
it is only the source-term computation kernel that is parallelized using OpenCL in that work. In our present work, we treat 
the significantly more common, {\em homogeneous} PDE i.e. the no source-term case, and parallelize {\em all} the compute kernels using OpenCL.} 
detailed evaluations of OpenCL for scientific computing on a wide variety of CPUs and GPUs. Lastly, all the computations performed 
here are in {\em double-precision} floating-point accuracy. 

This article is organized as follows: In Section 2, we provide a very brief introduction to the OpenCL parallel programming framework. 
In Section 3, we introduce our Teukolsky code, the relevant background gravitational physics and the numerical method used by the code. Next, we 
emphasize aspects of OpenCL and the compute hardware relevant to our implementation in Section 4. In Sections 5 \& 6, we present the overall 
performance results from our OpenCL-based parallelization efforts. Finally in Section 7, we summarize our work and make some conclusive 
remarks.

\section{OpenCL}
As mentioned already, OpenCL is a new framework for programming across a wide variety of computer hardware architectures 
(CPU, GPU, Cell BE, etc). In essence, OpenCL incorporates the changes necessary to the programming language C, that allow for 
parallel computing on all these different processor architectures. In addition, it establishes numerical precision requirements 
to provide mathematical consistency across the different hardware and vendors -- a matter that is of significant importance to 
the scientific computing community. Computational scientists would need to rewrite the performance intensive routines in their 
codes as OpenCL {\em kernels} that would be executed on the compute hardware. The OpenCL API provides the programmer various 
functions from locating the OpenCL enabled hardware on a system to compiling, submitting, queuing and synchronizing the compute 
kernels on the hardware. Finally, it is the OpenCL runtime that actually executes the kernels and manages the needed data transfers 
in an efficient manner. As mentioned already, most vendors have released an OpenCL implementation for their own hardware. As of 
the writing of the document, AMD and Nvidia have OpenCL freely available for their GPUs. IBM has also {\em beta} released OpenCL for the 
Cell BE and their Power line of multi-core processors.

OpenCL is of tremendous value to the scientific community because it is open, royalty-free and 
vendor- and platform- neutral. It delivers a high degree of {\em portability} across all major forms of current 
and future compute hardware. OpenCL allows us to run {\em identical} source-code on current (and 
future) multi-core CPUs, many-core GPUs and even {\em hybrid} processors such as the Cell BE and AMD Fusion, 
taking advantage of the different flavor of parallelism that they offer.  

\section{Numerical Relativity}
\label{}
In multiple earlier works~\cite{klpa,mystuff6,mystuff7} our time-domain  
Teukolsky equation solver is described in detail. Therefore, we simply reproduce some of that content below for completeness 
with minimal alterations.

Several gravitational wave observatories~\cite{ligo} are currently being built all over the world: LIGO in the United 
States, GEO/Virgo in Europe and TAMA in Japan. These observatories will open a new window onto the Universe by 
enabling scientists to make astronomical observations using a completely new medium -- gravitational waves (GWs), 
as opposed to electromagnetic waves (light). These GWs were predicted by Einstein's Relativity theory, but have 
not been directly observed because the required experimental sensitivity was simply not advanced enough, until 
very recently.

Numerical Relativity is an area of computational science that emphasizes the detailed modeling of 
strong sources of GWs -- collisions of compact astrophysical objects, such as neutron stars and black holes. Thus, NR plays an 
extremely important role in the area of GW astronomy and gravitational physics, in general. Moreover, the NR 
community has also contributed to the broader computational science community by developing an open-source, modular, 
parallel computing infrastructure called {\it Cactus}~\cite{cactus}. For the purposes of GW data analysis (detection and 
parameter estimation), it is critical to have a highly-accurate template bank of theoretical {\em waveforms}. Because of the 
degree of accuracy necessary and the large number of templates required, it is important to develop efficient computational 
methods for generating these theoretical waveforms. This motivates us to explore parallel computing frameworks like OpenCL and 
cutting-edge compute hardware like GPUs for NR.

The specific NR application we have chosen for consideration in this work is one that evolves the perturbations 
of a rotating (Kerr) black hole i.e. solves the Teukolsky equation in the time-domain~\cite{klpa,mystuff6,mystuff7}. 
This equation is essentially a linear wave-equation in Kerr space-time geometry. The next two subsections provide more 
detailed information on this equation and the associated numerical solver code.

\subsection{Teukolsky Equation}
The Teukolsky master equation describes scalar, vector and tensor field perturbations in the space-time of
Kerr black holes~\cite{eqn}. In Boyer-Lindquist coordinates, this equation takes the form
\begin{eqnarray}
\label{teuk0}
&&
-\left[\frac{(r^2 + a^2)^2 }{\Delta}-a^2\sin^2\theta\right]
         \partial_{tt}\Psi
-\frac{4 M a r}{\Delta}
         \partial_{t\phi}\Psi \nonumber \\
&&- 2s\left[r-\frac{M(r^2-a^2)}{\Delta}+ia\cos\theta\right]
         \partial_t\Psi\nonumber\\  
&&
+\,\Delta^{-s}\partial_r\left(\Delta^{s+1}\partial_r\Psi\right)
+\frac{1}{\sin\theta}\partial_\theta
\left(\sin\theta\partial_\theta\Psi\right)+\nonumber\\
&& \left[\frac{1}{\sin^2\theta}-\frac{a^2}{\Delta}\right] 
\partial_{\phi\phi}\Psi +\, 2s \left[\frac{a (r-M)}{\Delta} 
+ \frac{i \cos\theta}{\sin^2\theta}\right] \partial_\phi\Psi  \nonumber\\
&&- \left(s^2 \cot^2\theta - s \right) \Psi = 0  ,
\end{eqnarray}
where $M$ is the mass of the black hole, $a$ its angular momentum per unit mass, $\Delta = r^2 - 2 M r + a^2$ and 
$s$ is the ``spin weight'' of the field. The $s = 0$ versions of these equations describe the radiative degrees 
of freedom of a simple scalar field, and are the equations of interest in this work. As mentioned previously, this equation 
is an example of linear, hyperbolic, homogeneous (3+1)D PDEs which are quite common in several areas of science and 
engineering, and can be solved numerically using a variety of finite-difference schemes.  

\subsection{Teukolsky Code}
To solve Eq.\ (\ref{teuk0}) numerically in time-domain we take the approach first introduced by Krivan et al. in 
Ref.~\cite{klpa}. First, we make use of Kerr spacetime's axisymmetry and factor out the $\phi$-dependence of the 
Eq.\ (\ref{teuk0}) by decomposing the solution $\Psi$ into azimuthal $m$-modes
\begin{eqnarray}
\Psi(t,r,\theta,\phi) & = & \sum_m e^{im\phi}r^3\Phi_m(t,r,\theta)\;.
\end{eqnarray} 
In this manner the Eq.\ (\ref{teuk0}) is reduced to a linear system of decoupled (2+1)-dimensional hyperbolic PDEs. 
Then, we rewrite this system in first-order form, by introducing a new auxiliary ``momentum'' field variable, $\Pi$. 
And finally, we develop an explicit time-evolution numerical scheme for this first-order, linear PDE system using the 
well-known two-step, 2nd-order Lax-Wendroff, finite-difference method. Explicit details on this approach can be found in 
Ref.~\cite{klpa}. 

Each iteration to evolve the system above consists of two steps: In the first step, the solution vector 
$\mbox{\boldmath{$u$}}\equiv\{\Phi_R,\Phi_I,\Pi_R,\Pi_I\}$ between grid points is obtained from
\begin{eqnarray}
\label{lw1}
\mbox{\boldmath{$u$}}^{n+1/2}_{i+1/2} &=& 
\frac{1}{2} \left( \mbox{\boldmath{$u$}}^{n}_{i+1}
                  +\mbox{\boldmath{$u$}}^{n}_{i}\right)
- \\
&  &\frac{\delta t}{2}\,\left[\frac{1}{\delta r^*} \mbox{\boldmath{$D$}}^{n}_{i+1/2}
  \left(\mbox{\boldmath{$u$}}^{n}_{i+1}
                  -\mbox{\boldmath{$u$}}^{n}_{i}\right)
- \mbox{\boldmath{$S$}}^{n}_{i+1/2} \right] \; .\nonumber
\end{eqnarray}
This is used to compute the solution vector at the next time step,
\begin{equation}
\mbox{\boldmath{$u$}}^{n+1}_{i} = 
\mbox{\boldmath{$u$}}^{n}_{i}
- \delta t\, \left[\frac{1}{\delta r^*} \mbox{\boldmath{$D$}}^{n+1/2}_{i}
  \left(\mbox{\boldmath{$u$}}^{n+1/2}_{i+1/2}
                  -\mbox{\boldmath{$u$}}^{n+1/2}_{i-1/2}\right)
- \mbox{\boldmath{$S$}}^{n+1/2}_{i} \right] \, .
\label{lw2}
\end{equation}
The angular subscripts are dropped in the above equation for clarity. All angular derivatives are computed using 2nd-order, 
centered finite difference expressions. Explicit forms for the matrices {\boldmath{$D$}} and {\boldmath{$S$}} can be easily 
found in the relevant literature~\cite{klpa}.

Symmetries of the spheroidal harmonics are used to determine the angular boundary conditions: 
For even $|m|$ modes, we have $\partial_\theta\Phi =0$ at $\theta = 0,\pi$ while $\Phi =0$ at $\theta = 0,\pi$ for 
modes of odd $|m|$. We set $\Phi$ and $\Pi$ to zero on the inner and outer radial boundaries.

\section{OpenCL Implementation}
\label{}
The first task in our work is to isolate the most compute intensive portions of our Teukolsky equation solver code. 
Upon performing a basic profiling of our code using the GNU profiler {\bf gprof}, we learn that computing the 
``right-hand-sides'' of the Lax-Wendroff steps i.e. the quantities within the square-brackets of Eqs.\ (\ref{lw1}) 
and (\ref{lw2}), take nearly {\bf 75\%} of the application's overall runtime. We anticipate that this observation 
is fairly typical for codes of this type. Thus, it is natural to consider accelerating this ``right-hand-side'' 
computation using data-parallelization on the many cores of the GPU. 

A data-parallel model is relatively straightforward to implement in a code like ours. We simply perform a domain 
decomposition of our finite-difference numerical grid and allocate the different parts of the grid to different cores. 
More specifically, on the GPU, each thread computes the right-hand-side for a {\em single} pair of $r$ and $\theta$ grid 
values. In addition, it is necessary to establish the appropriate data communication between the GPU cores and the 
remaining code that is executing on the CPU -- we use {\bf  clEnqueueReadBuffer, clEnqueueWriteBuffer} instructions 
to transfer data back-and-forth from main memory and we only use {\em global memory} on the GPU to simplify communication 
between the GPU cores. We make this simplification (only making use of global memory) for the stated goal of keeping 
the code's {\em portability} intact, even if it impacts performance to some extent~\footnote{Here, we very briefly document 
our experiences with two GPU-specific optimization techniques. Besides global memory, there are many other kinds of memory 
available on a GPU. Global memory is large, but is rather slow to access. On the other hand, {\em shared memory} is small, 
but very fast to access. If there is a lot of reuse of a particular data, it makes sense to store it in shared memory, thus 
enabling much faster access to this data. In our OpenCL Teukolsky code, there is not enough reuse of any memory data. Hence, 
shared memory optimizations yield only few percent of performance improvement. Another optimization technique that is 
frequently used in GPUs is {\em memory coalescing}. To achieve the peak memory bandwidth, global memory accesses needs to 
be coalesced. This means that memory locations accessed by the kernels need to be from contiguous memory locations. We note 
that memory accesses in our OpenCL code are fairly regular and therefore memory accesses are already coalesced.}. 

Unfortunately, this naive approach yields a {\em negligible} performance gain on the GPU. The reason is that although the 
right-hand-side computation is accelerated due to the use of the many-cores of the GPU, the time it takes to bring that 
data back-and-forth from main memory so that the remaining computation can resume on the CPU, is large enough that no 
overall gain in performance is perceived! This outcome is simply due to the poor bandwidth of the system's PCI bus on which 
the GPU is located. To address this issue, we port {\em all} the Lax-Wendroff related compute routines (such as the 
computation of the evolved fields half-way between grid points, the boundary condition imposition, updating of the fields 
using the right-hand-side data) as separate kernels onto the GPU. In this manner, no communication is necessary with the 
rest of the computer system and we overcome the challenge mentioned above. It is worth noting that some of these routines 
are perhaps not ideal for execution on the GPU (for example, some don't quite have a high enough {\em arithmetic intensity} 
that is essential to obtain high performance from the GPU architecture) but we still port these over for execution on the 
GPU regardless, simply because our goal is to minimize data transfer back-and-forth from main memory. This requires a significant 
amount of additional effort -- but one that pays off well eventually (as seen in the following section). 

The main limitation that we introduce with this approach of running the entire computation on the GPU is that we need 
to able to fit the entire memory requirements of the code within the GPU video memory. Given that current high-end GPU 
offerings support only a few GBs of memory, this can be challenging for most NR codes, especially those in (3+1)D. However, a 
compute cluster with multiple GPUs per node, could perhaps overcome this serious limitation. Most NR codes are message-passing 
(MPI) parallelized for cluster execution already using domain-decomposition. It should not be too difficult to extend that 
approach to a cluster with GPUs as the main compute devices. For the performance tests in our current work, we simply use grid 
sizes that maximize the use of video memory on the (single) GPUs we consider.

Below, we depict a sample kernel from the OpenCL code. This kernel computes the evolved fields half-way between the grid points 
using a simple averaging process. The array variables are defined as follows: {\bf qre} and {\bf qim} are the real and imaginary 
parts of the solution $\Phi$, while {\bf pre} and {\bf pim} are the real and imaginary parts of the ``momentum'' $\Pi$. The integers 
{\bf a} and {\bf b} label the array indices that are relevant to the $r$ and $\theta$ grid values involved in the kernel.

{\small
\begin{verbatim}

#pragma OPENCL EXTENSION cl_khr_fp64: enable
#define idx(b,a) (N*(b)+(a))

/* --------------------------------------- */

__kernel void kernel_average1 (
__global double* qre_h, __global double* qim_h, 
__global double* pre_h, __global double* pim_h, 
__global double* qre, __global double* qim, 
__global double* pre, __global double* pim)
{
 int b; int a;  

 b = get_global_id(0);
 a = get_global_id(1);

 if (a >= 1 && a < N)
  {
  qre_h[idx(b,a)] = 0.5 * (qre[idx(b,a)] 
                          + qre[idx(b,a-1)]);
  qim_h[idx(b,a)] = 0.5 * (qim[idx(b,a)] 
                          + qim[idx(b,a-1)]);
  pre_h[idx(b,a)] = 0.5 * (pre[idx(b,a)] 
                          + pre[idx(b,a-1)]);
  pim_h[idx(b,a)] = 0.5 * (pim[idx(b,a)] 
                          + pim[idx(b,a-1)]);
  }
}

/* --------------------------------------- */

\end{verbatim}
}

In summary, it is worth pointing out that this OpenCL implementation of our Teukolsky solver code is fairly straightforward. 
It should also be mentioned that we do not make a serious attempt to hand-tune the codes to tailor them for each architecture, 
in order to obtain maximal performance. As stated earlier, one of our goals is to keep the code highly portable, because we 
aim to run the exact same code on both GPUs and CPUs. The only variable that we tune (through simple experimentation) in order 
to obtain maximum performance for each architecture is the {\bf local\_work\_size}. 

\section{Performance Results}
\label{}
In this section, we report on the final results from our OpenCL implementation as outlined in the previous section.
We begin by documenting the details of the compute hardware we use to perform the OpenCL code testing. We use three 
types of CPUs: Intel ``Harpertown'' Xeon (while this CPU is a bit dated, it is still one of the common ones found 
in compute clusters at the present time, and therefore is still of some interest); Intel ``Nehalem'' Xeon (the most 
current Intel offering); IBM Power7 (the most current processor from IBM's Power line -- one that will be used in NSF's 
Blue Waters {\em petascale} system in the very near future, and therefore of particular interest). We use two types of 
high-end GPUs: Nvidia's ``Fermi'' C2050 (most current CUDA Tesla GPU offering from Nvidia with 3GB memory); AMD 
``Radeon'' 5870 (most current Stream GPU offering from AMD with 1GB memory). In addition, we also include results 
from a traditional, message-passing (MPI) based, 10-node compute cluster built using 80 Intel ``Harpertown'' Xeon cores and 
high-speed Infiniband DDR interconnects. Detailed specifications (GHz, cores, etc.) of these processors are available in Table 1. 
The systems supporting this compute hardware are either running Mac OS X (the Nehalem and Harpertown CPUs) or a Linux distribution 
(everything else) as the primary operating system. Finally, the most current OpenCL distribution available publicly at the 
time of performing this work, is installed on these systems.

\begin{table}
 \caption{\it This table depicts the relative values for performance for several variants of current generation 
CPUs and GPUs. The baseline system here is an Intel ``Harpertown'' Xeon, 8-core, 2.8 GHz CPU running our OpenCL code. 
These values are based on the overall runtimes of our Teukolsky code on these different systems.}
  \centering
     \begin{tabular}{|c|c|c|c|c|c|} \hline
   {\bf Name} & {\bf Type} & {\bf GHz} & {\bf Cores} & {\bf Code} & {\bf Perf.}\cr
    \hline \hline
    {\bf Nehalem} & CPU &   2.3    &  8    & OCL & {\bf 1.4x} \cr \hline
    {\bf Power7}  & CPU &   3.0    &  8    & OCL & {\bf 1.4x} \cr \hline
    {\bf Fermi}   & GPU &   1.2    &  448  & OCL & {\bf 9.9x} \cr \hline
    {\bf Radeon}  & GPU &   0.9    &  1600  & OCL & {\bf 9.5x} \cr \hline
    {\bf Cluster} & CPU &   2.5    &  80   & MPI & {\bf 9.6x} \cr \hline
   \end{tabular}
\label{comp}
\end{table}

The last column of Table 1 depicts our final performance numbers\footnote{When compared with 
the serial (single-core) code, the OpenCL version yields a performance gain of a factor of 2.5 on 8 CPU cores. 
This modest scaling is fairly typical of NR codes like these. This is mainly due to the fact such codes are 
severely {\em memory bandwidth} limited and therefore, it is the slow access to system memory that strongly limits 
their overall performance.}. It is clearly evident that the GPUs yield an {\em order-of-magnitude} gain in overall performance 
over multi-core CPUs (even in a {\em dual} i.e. 8-core configuration). Recall, that the entire computation performed here
is in the context of the {\em double-precision} floating-point accuracy. What is also interesting is that this outcome is 
largely vendor and architecture {\em independent}. In other words, both Nvidia and AMD GPUs provide near identical gains over 
both Intel and IBM CPUs. In fact, all the CPUs considered in our study performed comparably and so did the 
GPUs, regardless of vendor or architecture. Note that in particular, Nehalem and Power7 processor architectures 
are significantly different (for example, one is x86 while the other is PPC), but our OpenCL code yields near 
identical performance on both.  

What is also striking is that a (single) GPU performs comparable to a small-sized CPU cluster! In particular,  
we observe that a high-end Nvidia or AMD GPU can match the performance of an 80-core (10-node) Intel Xeon CPU cluster 
with high-speed interconnects. This suggests that a multi-GPU desktop system can potentially replace common small/mid 
sized CPU clusters (upto several 100 cores); which would yield significant savings in physical space, 
procurement costs, power consumption and a major reduction in failure-rate.

Another observation worth making from our results above is that the GPUs also outperform the CPUs in metrics such as 
{\em performance-per-dollar} and {\em performance-per-watt} by an {\em order-of-magnitude}. In particular, between the 
Fermi and the Radeon, since they both exhibit the same level of performance, but the Radeon is over five (5) 
times lower in cost and also slightly more power efficient, it is the most cost-effective compute hardware
amongst the ones we consider in this work. Although, it should also be noted that the Radeon only has 1GB of memory, 
compared to the Fermi that is equipped with with a substantially higher, 3GBs. 

Finally, it is worth emphasizing that we run {\em identical} source-code on all the hardware we consider in this work.
That is a highly non-trivial benefit offered by the OpenCL framework, promising tremendous savings in code-development 
efforts to the computational scientist. As mentioned earlier, the goal of our present work is to evaluate OpenCL not 
only from the viewpoint of performance, but also {\em portability}. The fact that we are able to run the same source-code 
on processor architectures as different as CPUs and GPUs, and yet achieve high performance, speaks very highly of this 
technology and its potential in scientific computing.  

\section{Other Related Work}
\label{}
Since OpenCL is a relatively new framework, currently there are not many published results based on it in the relevant 
literature. However, there are a few closely related works that we mention in this section. Karimi et al.~\cite{karimi}  
perform a performance comparison of CUDA and OpenCL on Nvidia GPUs and find that CUDA is faster on the kernels they ran. 
Waage~\cite{waage} uses OpenCL to accelerate image filtering and obtains comparable results to a CUDA implementation. Zhang
et al.~\cite{zhang} use OpenCL to accelerate CT reconstruction and image recognition. Brown et al.~\cite{brown} develop an 
OpenCL and CUDA implementation of molecular dynamics software LAMMPS and find that OpenCL is somewhat slower than CUDA.    

\section{Conclusions}
\label{}
The main goal of this work is to evaluate a new parallel code development framework, OpenCL, for scientific computation. 
OpenCL is hardware and platform neutral and yet able to deliver strong performance i.e. it delivers {\em portability} and 
{\em high performance} on all modern many-core GPUs and multi-core CPUs. This makes OpenCL potentially very attractive 
to the scientific computation community.

In this work, we perform one of the first careful evaluations of OpenCL for scientific computing on a wide variety of 
currently available CPUs and GPUs. In particular, we take a sample application from the Numerical Relativity community -- 
a time-domain, linear, finite-difference, hyperbolic PDE solver -- and implement its entire computation as parallel OpenCL 
kernels. We describe the parallelization approach taken and also the relevant important aspects of the considered compute 
hardware in some detail. 

Our results suggest that it is relatively straightforward to obtain {\em order-of-magnitude} gains in overall application 
performance by using current Nvidia or AMD GPUs over Intel or IBM CPUs. In addition, this outcome is largely vendor and 
architecture independent. Moreover, the OpenCL source-code is identical for all these CPUs and GPUs, which is a non-trivial 
benefit; it promises significant savings in parallel code-development and optimization efforts. 

\section* {Acknowledgements}
We would like to thank Geoff Cowles, Glenn Volkema and Robert Fisher for their assistance with this work throughout, many 
helpful discussions and also for providing useful feedback on this manuscript. GK and RG would also like to acknowledge 
research support from the National Science Foundation (NSF grant numbers: PHY-0902026, CNS-0959382, PHY-1016906, PHY-1135664).

\end{document}